\documentclass[conference]{IEEEtran}
\ifCLASSINFOpdf
\else
\fi
\usepackage{amsthm}
\usepackage{makeidx}
\usepackage{examples}
\usepackage[pdftex]{graphicx}
\usepackage{amssymb}
\usepackage{graphicx}
\usepackage{algorithm}
\usepackage{epstopdf}
\usepackage{color}
\usepackage{url}
\graphicspath{{../eps/}{../ps/}}
\usepackage{psfrag}
\usepackage{algorithm}
\usepackage{algorithmic}
\usepackage{amsmath}
\font\msbm=msbm10 at 9pt
\newcommand{\ZZ}{\mbox{\msbm Z}}

\newcommand{\CC}{\mbox{\msbm C}}
\newcommand{\FF}{\mbox{\msbm F}}

\def \C {{\CC}}

\def \F {{\FF}}

\newtheorem{theorem}{Theorem}

\newtheorem{remark}[theorem]{Remark}
\newtheorem{example}[theorem]{Example}
\newtheorem{defenition}[theorem]{Defenition}

\begin{document}
%
\title{Enumerating Some Fractional Repetition Codes}

\author{
\IEEEauthorblockN{Srijan Anil}
\IEEEauthorblockA{Sapient-Nitro\\
Gurgaon, Delhi, India\\
Email:sanil@sapient.com}
\and
\IEEEauthorblockN{Manish K. Gupta}
\IEEEauthorblockA{Laboratory of Natural Information Processing\\
Dhirubhai Ambani Institute of Information\\ and Communication Technology\\
Gandhinagar, Gujarat, 382007 India\\
Email: mankg@computer.org}
\and
\IEEEauthorblockN{T. Aaron Gulliver}
\IEEEauthorblockA{Department of Electrical and Computer Engineering\\
University of Victoria
Victoria, BC, V8W 3P6 Canada\\
Email: agullive@ece.uvic.ca}
}


%


\maketitle

\begin{abstract}
In a distributed storage systems (DSS), regenerating codes are used to optimize bandwidth in the repair process of a failed node.
To optimize other DSS parameters such as computation and disk I/O, Distributed Replication-based Simple Storage (Dress) Codes consisting
of an inner Fractional Repetition (FR) code and an outer MDS code are commonly used.
Thus constructing FR codes is an important research problem, and several constructions using graphs and designs have been proposed.
In this paper, we present an algorithm for constructing the node-packet distribution matrix of FR codes and thus enumerate some FR codes
up to a given number of nodes $n$. We also present algorithms for constructing regular graphs which give rise to FR codes.
\end{abstract}


%
\IEEEpeerreviewmaketitle

\section{Introduction}
The emerging era of cloud computing poses new challenges for researchers to provide reliable and secure data storage.
Practical systems for distributed storage include the Hadoop based system \cite{XorbasVLDB} used in Facebook and Windows Azure storage \cite{Huang:2012:ECW:2342821.2342823}.
In these distributed storage systems (DSSs), data is stored on $n$ unreliable nodes.
Reliability is provided either by replicating the data or using erasure MDS (Maximum Distance Separable) codes.
Both of these schemes have drawbacks either in terms of bandwidth, complexity or disk I/O.
To overcome these limitations, regenerating codes were introduced by Dimakis et al. \cite{dgwr7},
and subsequently studied by many researchers ~\cite{dgwr7,survey,dress11,DBLP:journals/corr/abs-1211-1932,2013arXiv1302.0744K,XorbasVLDB,Gupta:arXiv1302.3681}.
A node failure in such systems can be handled by regenerating the data stored on that node using its peers.
This regeneration can be functional or exact.
Functional repair allows restoration of the data such that a stored file can be retrieved by contacting any $k$ out of $n$ nodes, where $k < n$.
Exact repair allows for the creation of a replica of the data previously stored on the node~\cite{survey,rr10}.
Regenerating codes are specified by the parameters $\{ [n, k, d], [\alpha,\beta, B] \}$,
where $n$ is the number of nodes, $k$ is the number of nodes that need to be contacted to recover a file $B$, and
$d$ is the repair degree (the number of nodes that must be contacted to regenerate data in case of a node failure).
The capacity of a node is given by $\alpha$, and the repair bandwidth for each of the $d$ nodes is $\beta$,
so the total repair bandwidth is $d\beta$ ~\cite{rr10}.

The tradeoff in Regenerating codes between the storage capacity and repair bandwidth have given rise to two new classes of codes,
namely Minimum Storage Regenerating (MSR) codes and Minimum Bandwidth Regenerating (MBR) codes.
MBR codes employ \emph{exact} and \emph{uncoded} data repair.
Uncoded repair means that a particular set of $d$ nodes, as listed in the Repair Table of the node, are contacted and one data packet is
downloaded from each, thus reducing the repair complexity.
MBR codes are formed by the concatenation of an outer MDS code and an inner Fractional Repetition (FR) code.
The MDS code maintains the MDS property of the DSS, while the Fractional Repetition codes allow for an uncoded repair process.
These concatenated codes are known as DRESS codes (Distributed Replication-based Exact Simple Storage) codes~\cite{rr10,dress11}.
Many constructions of Fractional Repetition Codes (and hence DRESS codes),
are known based on bipartite graph \cite{DBLP:journals/corr/abs-1102-3493}, resolvable designs \cite{DBLP:journals/corr/abs-1210-2110}, regular graphs \cite{rr10,Wangwang12} and other structures~\cite{Gupta:arXiv1302.3681}.

In this paper, an algorithm for the construction of Fractional Repetition (FR) codes
is presented which is based on the incidence matrix of the node-packet distribution.
Algorithms are also given for the construction of regular graphs.
The rest of the paper is organized as follows.
In Section 2, the basics of FR codes and the incidence matrix of the node-packet distribution are given.
Section 3 presents an algorithm  for the construction of the $n \times \theta$ incidence matrix of the node-packet distribution of an FR code.
Algorithms for constructing regular graphs and hence FR codes for $n=\theta$ are presented in Section 4.
Finally, Section 5 concludes the paper with some general remarks.


\section{Background}
\begin{figure}
\centering
\includegraphics[scale=0.65]{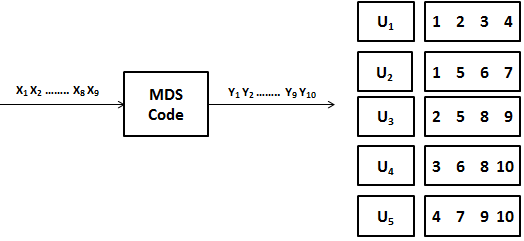}
\caption{A DRESS code consisting of an inner fractional repetition code $\C$ having $n=5$ nodes, $\theta=10$ packets,
replication factor $\rho=2$, and repair degree $d=4$, and an outer MDS code.}
\label{dress}
\end{figure}

Distributed Replication-based Simple Storage (DRESS) codes consist of an inner Fractional Repetition (FR) code and an outer MDS code,
as shown in Figure \ref{dress}.
FR codes are formally defined in Definition \ref{defFRR}.

\begin{defenition}(Fractional Repetition Code):
A Fractional Repetition (FR) code $\C$, with repetition degree $\rho$, for an $(n,k,d)$ DSS, is a collection $\C$ of $n$ subsets
$U_1,U_2, \ldots ,U_n$ of a set $\Omega = \{1,\ldots,\theta \}$, each having size $d$, i.e, $|U_i|=d$, satisfying the condition that each element of
$\Omega$ belongs to exactly $\rho$ sets in the collection.
The code is denoted by $\C: (n, \theta, d, \rho)$, and the parameters of $\C$ are related by $n d = \rho \theta$.
\label{defFRR}
\end{defenition}

\begin{example}
Figure \ref{dress} gives an example of FR code.
Suppose there are 9 packets from $\F_q$ (a finite field with $q$ elements), and 5 storage nodes.
Using an MDS code, the 9 packets are first encoded into 10 packets such that the last packet is the parity packet.
Next all 10 packets are replicated twice ($\rho=2$), on the 5 nodes according to the arrangement of the FR code $\C: (5, 10, 4, 2)$ in the figure.
This code can tolerate 1 failure and the data can be recovered by contacting 4 nodes, hence the repair degree is 4.
\end{example}

\begin{remark}
An FR code $\C: (n, \theta, d, \rho)$ can also be characterized by a node-packet distribution incidence matrix $M$ of size $n \times \theta$
with row weight $d$ and column weight $\rho$.
For example, the incidence matrix for the FR code $\C: (5, 10, 4, 2)$ shown in Figure \ref{dress} is given by Table \ref{nodep}.
The row weight is 4 and the column weight is 2.

\begin{table}
\caption{Node-Packet Distribution Incidence Matrix $M$ of Size $5 \times 10$ for the FR Code $\C: (5, 10, 4, 2)$ Shown in Figure \ref{dress}}
ÊÊÊÊ\begin{tabular}{|c||c|c|c|c|c|c|c|c|c|c|}
\hline
ÊÊÊÊÊÊÊÊNode/Packets & $P_1$ & $P_2$ & $P_3$ & $P_4$ & $P_5$ & $P_6$ & $P_7$ & $P_8$ & $P_9$ & $P_{10}$ \\\hline\hline
ÊÊÊÊÊÊÊÊ$U_1$Ê & 1ÊÊ & 1ÊÊ & 1ÊÊ & 1ÊÊ & 0ÊÊ & 0ÊÊ & 0ÊÊ & 0ÊÊ & 0ÊÊ & 0ÊÊÊÊÊ \\
ÊÊÊÊÊÊÊÊ$U_2$ÊÊ& 1ÊÊ & 0ÊÊ & 0ÊÊ & 0ÊÊ & 1ÊÊ & 1ÊÊ & 1ÊÊ & 0ÊÊ & 0ÊÊ & 0ÊÊÊÊÊ \\
ÊÊÊÊÊÊÊÊ$U_3$Ê & 0ÊÊ & 1ÊÊ & 0ÊÊ & 0ÊÊ & 1ÊÊ & 0ÊÊ & 0ÊÊ & 1ÊÊ & 1ÊÊ & 0ÊÊÊÊÊ \\
ÊÊÊÊÊÊÊÊ$U_4$ÊÊ& 0ÊÊ & 0ÊÊ & 1ÊÊ & 0ÊÊ & 0ÊÊ & 1ÊÊ & 0ÊÊ & 1ÊÊ & 0ÊÊ & 1ÊÊÊÊÊ \\
ÊÊÊÊÊÊÊÊ$U_5$ÊÊ& 0ÊÊ & 0ÊÊ & 0ÊÊ & 1ÊÊ & 0ÊÊ & 0ÊÊ & 1ÊÊ & 0ÊÊ & 1ÊÊ & 1ÊÊ\\
\hline
\end{tabular}
\label{nodep}
\end{table}
\end{remark}
\subsection{Equivalence of Fractional Repetition codes}
Two Fractional Repetition codes $\C_1: (n_1, \theta_1, d_1, \rho_1)$  and $\C_2: (n_2, \theta_2, d_2, \rho_2)$ are said to be equivalent if
\begin{enumerate}
\item The number of nodes and the number of packets in the system are same, i.e., $n_1=n_2$ and $\theta_1=\theta_2$. Hence the dimension of the corresponding incidence matrices is the same, i.e., $n_1 \times \theta_1 = n_2 \times \theta_2$.
\item The repair degree and the replication factor are the same, i.e., $d_1= d_2$ and $\rho_1 = \rho_2$. Hence the corresponding incidence matrices have the same row weight $d$ and column weigh $\rho$.
\item The same packet distribution can be achieved by simply renaming the packets of one of the codes i.e., if the incidnece matrix of $\C_1$ can be obtained by applying permutations on the rows and columns of the incidence matrix of $\C_2$.
\end{enumerate}

\begin{remark}
An incidence matrix of dimension $n \times \theta$ defines an FR code with $n$ nodes and $\theta$ packets.
The repair degree is $d$, and the replication factor is $\rho$.
Now, taking the transpose of this matrix gives a matrix of dimension $\theta\times n$.
The weight of each row is now $\rho$, and the weight of each column is $d$.
This new matrix also satisfies the conditions for an FR code, and
corresponds to a code with $\theta$ nodes, $n$ packets, repair degree $\rho$, and replication factor $d$.
\end{remark}
\section{Enumeration of Fractional Repetition Codes using Incidence Matrices}
\begin{figure}
\centering
\includegraphics[scale=0.20]{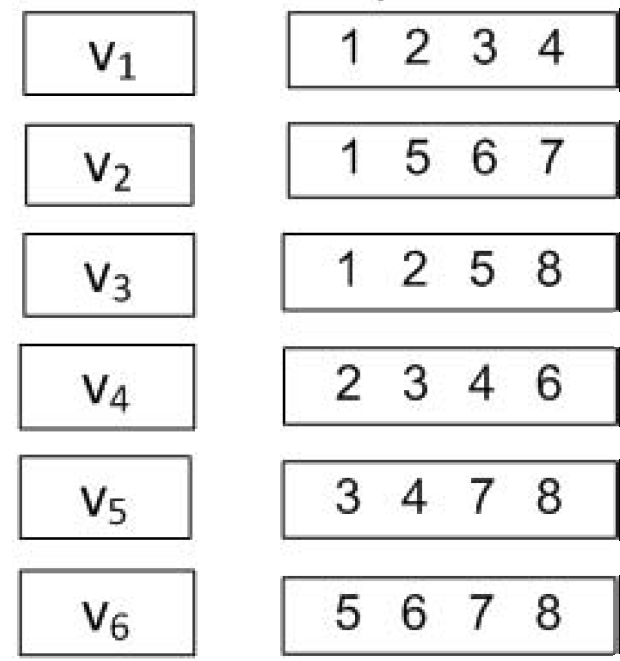}
\caption{The FR code $\C:(6,8,4,3)$ generated using the incidence matrix in Example 5.}
\label{genaloexample}
\end{figure}

To enumerate the FR codes for a given $n$, the replication factor $\rho$ can be varied in the range $2 \leq \rho \leq n-1$,
and the repair degree $d$ in the range $2  \leq d \leq n-1$.
In each case, $\theta$ can be determined using $n d = \rho \theta$ and the corresponding incidence matrix $M$ of size $n \times \theta$ can be filled
such that the weight of each row is $d$ and the weight of each column is $\rho$ to obtain an FR code.
Algorithm \ref{algonbytheta} is given below to fill the incidence matrix with $1's$ and $0's$.
Table \ref{possibleFRcodes} summarizes the number of possible FR codes up to length $n=10$.
For larger $n$, the data can be obtained from \url{http://www.ece.uvic.ca/~agullive/manish/List.html}.
\begin{algorithm}
\caption{Generate a node-packet distribution incidence matrix $M$ of size $n \times \theta$}
\begin{algorithmic}
\REQUIRE $n, d , \theta , \rho$ and an all zero matrix $M$ of size $n \times \theta$
\ENSURE M$_{n \times \theta}$\; \mbox{such that} $\;weight(row[M])=d$ \newline \mbox{and} $\; weight(column[M])= \rho$
\STATE$1:$ Place a 1 in $d$ positions of the $1^{st}$ row from left to right starting from $m_{11}$ and move to the $2^{nd}$ row.
\STATE$2:$ In the row, place a 1 in  the first column $j, 2 \leq j \leq \theta$ for which the column weight is $< \rho$.
\STATE$3:$ Compute the weight of all consecutive columns from $j+1$ to $\theta$. If the minimum weight of these columns is the same, go to Step 4,
otherwise place 1's in increasing order of weight until $weight(row) = d$ or the last column is reached. Go to Step 6
\STATE$4:$ Traversing rows from the top, identify the first row having an entry 1 which corresponds to a 1 in the $j^{th}$ column (determined in Step 2),
in the current row.
\STATE$5:$ Traversing consecutive columns from $j+1$ to $\theta$ in the current row, place a 1 in the column for which a 0 first occurs
in the row identified in Step 4.
\STATE$6:$ If $weight(row) < d$, go to Step 2 otherwise move to Step 7
\STATE$7:$ If a next row exists, move to that row and go to Step 2, otherwise Stop.
\end{algorithmic}
\label{algonbytheta}
\end{algorithm}
\begin{example}
For  $n=6, d= 4, \theta =8$ and $\rho=3$, Algorithm \ref{algonbytheta} gives the following incidence matrix
\[
M_{6 \times 8} =
\begin{bmatrix}1 & 1 & 1 & 1 & 0 & 0 & 0 & 0\\
1 & 0 & 0 & 0 & 1 & 1 & 1 & 0\\
1 & 1 & 0 & 0 & 1 & 0 & 0 & 1\\
0 & 1 & 1 & 1 & 0 & 1 & 0 & 0\\
0 & 0 & 1 & 1 & 0 & 0 & 1 & 1\\
0 & 0 & 0 & 0 & 1 & 1 & 1 & 1
\end{bmatrix}.
\]
This matrix gives the FR code $\C:(6,8,4,3)$ as shown in Figure \ref{genaloexample}.
\end{example}

\begin{table}
\begin{center}
\caption{The Number of Possible FR Codes for $n=3$ to 10 Nodes}
\begin{tabular}{|c|c|}
\hline
Number of Nodes ($n$) & Number of FR Codes \\\hline\hline
3 & 1 \\
4 & 3\\
5 & 4\\
6 & 10\\
7 & 8\\
8 & 16  \\
9 & 19 \\
10& 28 \\
\hline
\end{tabular}
\label{possibleFRcodes}
\end{center}
\end{table}

\section{Construction of Regular Graphs}
FR codes can be generated using a regular graphs of degree $d$ \cite{rr10,Wangwang12}.
Therefore, Algorithm \ref{regularalgo1} is presented for generating regular graphs. We also present
Algorithms \ref{regualgo2} and \ref{regualgo3}  for constructing regular graphs based on the approach of filling the incidence matrix to obtain an FR code.
To the best of our knowledge, this solution has not been reported in the vast literature on regular graphs.
An example is given for each algorithm.
The proposed algorithms are constrained to $nd \in 2\ZZ^{+}$ and $\rho=2$.
Note that a regular graph of degree $d$ is a graph where every vertex has the same degree $d$,
which is possible only for $nd \in 2\ZZ^{+}$.

\begin{algorithm}
\caption{Regular Graph for $nd \in 2\ZZ^{+}$, $\rho=2$ and $d<n-1$}
\begin{algorithmic}

\STATE$1:$ Divide the $n$ vertices into two set of vertices, $U \lbrace u_{1},u_{2}, \ldots ,u_{\lfloor\frac{n}{2}\rfloor}\rbrace$
and $V \lbrace v_{1}, v_{2}, \ldots, v_{\lceil\frac{n}{2}\rceil}\rbrace$
\STATE$2:$ Construct two cyclic graphs $G_{1}:(U, E_{1})$ and $G_{2}:(V, E_{2})$, with $G_{1}$ enclosing $G_{2}$\\
	\IF {$n$ is odd}
	\STATE Select two vertices $v_{i}$ and $v_{j}$ such that edge $\lbrace v_{i}, v_{j}\rbrace \notin  E_{2}$
	\STATE Add edge $\lbrace v_{i}, v_{j}\rbrace$
	\ENDIF				
\STATE Select vertices $u_{i} \in U$ and $v_{j} \in V$ such that $\deg(v_{j})\neq d$
\STATE \mbox{ }Add edge $\lbrace u_{i}, v_{j}\rbrace$
Repeat for vertex u$_{i}$ until $\deg(u_{i})=\lfloor\frac{d}{2}\rfloor$
\STATE Select vertices $u_{i}, u_{j} \in U$, such that edge $\lbrace u_{i}, u_{j} \rbrace \notin E_{1}$
\STATE \mbox{ }Add edge$\lbrace u_{i}, u_{j}\rbrace$
\STATE Repeat for vertex u$_{i}$ until $\deg(u_{i})=\lceil\frac{d}{2}\rceil$

Pick vertex $v_{i},v_{j} \in V$, such that edge $\lbrace v_{i}, v_{j}\rbrace \notin  E_{2}$
\STATE \mbox{ }Add edge $\lbrace v_{i},v_{j}\rbrace$
\STATE Repeat for vertex $v_{i}$ until $\deg(v_{i})=\lceil\frac{d}{2}\rceil$

\end{algorithmic}
\label{regularalgo1}
\end{algorithm}

\begin{figure}
\centering
\includegraphics[width=1.2in]{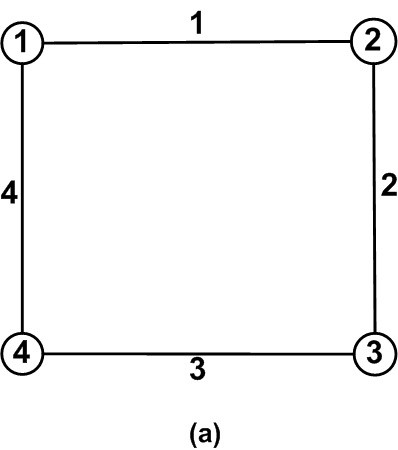}
\label{4 node}
\includegraphics[width=1.7in]{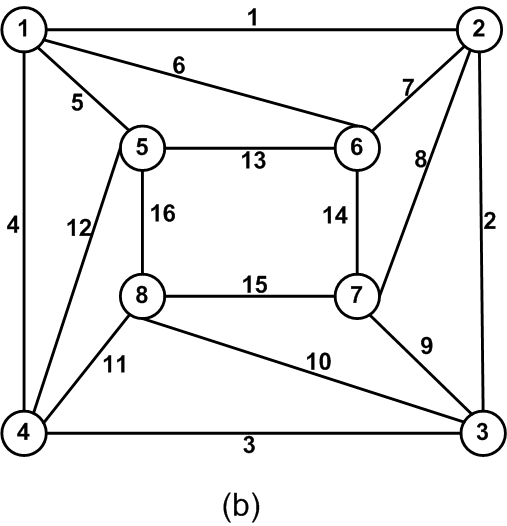}
\label{8 node}
\includegraphics[width=3.3in]{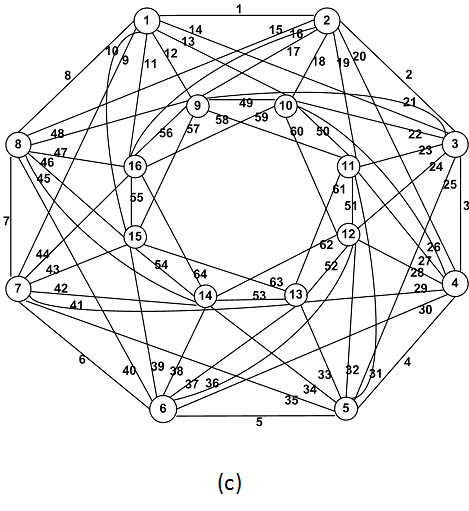}
\label{32 node}
\caption{A regular graph for $\rho=2$ and (a) $n=4, d=2$  (b) $n=8, d=4$   (c) $n=16, d=8$.
The vertices of the graphs are nodes, and the edges originating from them are the packets stored in those nodes.
Thus Algorithm \ref{regularalgo1} generates a $d$-regular graph which depicts the packet distribution among the nodes as shown in
Figure \ref{Configuration48nodes}}
\label{reggraph1fig}
\end{figure}

\begin{figure}
\centering
\includegraphics[width=3in]{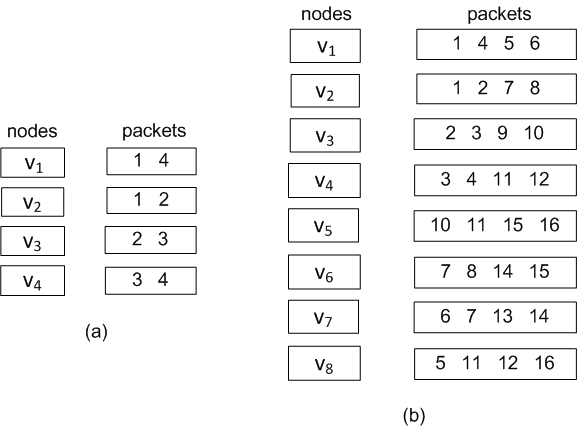}
\caption{Node configuration for (a)$n=4$, $\theta=4$, $d=2$, $\rho=2$ (b) $n=8$, $\theta=16$, $d=4$, $\rho=2$.
This distribution shows that any two nodes have either no packet or 1 packet in common.}
\label{Configuration48nodes}
\end{figure}

\begin{example}
Algorithm \ref{regularalgo1} can be used to generate a $d$-regular graph for $\rho=2$ and (a) $n=4$, $d=2$ (b) $n=8$, $d=4$ (c) $n=16$, $d=8$,
as shown in Figure \ref{reggraph1fig}, and a FR code as shown in Figure \ref{Configuration48nodes}.
\end{example}

An adjacency matrix is a matrix depicting the relationship between vertices, showing whether they are connected or not.
FR codes can be represented by graphs, where the vertices represent the nodes and the edges represent the packets.
These can be interchanged, thus making edges the nodes and vertices the packets.
Now for $nd$ even, and $\rho=2$, graphs can be represented by an adjacency matrix of dimensions $n \times n$.
This matrix acts as a basis for generating the incidence matrix of the graph.
The incidence matrix shows the packet distribution over the $n$ nodes.
We present two algorithms to generate the adjacency matrix for parameters $\rho$, $n$, $d$, $\theta$, with constraints $\rho$=2 and $nd \in 2\ZZ^{+}$,
where $n = \theta$, row weight $d$ and column weight $\rho$.
This provides an FR code with the same number of nodes and packets.

\begin{algorithm}
\caption{Adjacency Matrix $A$ of Size $n \times n$}
\begin{algorithmic}
\REQUIRE $n , d, \theta, \rho$ and a null matrix $A$ of size $n \times n$
\ENSURE  $A_{n \times n}$ such that $weight(row[A])=d$ \newline and $weight(column[A])= \rho$
\STATE$1:$ Set $a_{12}=1$ and fill the consecutive entries of first row with $(d-1)$ 1's from left to right
\STATE$2:$ Set the first column as the transpose of the first row
\STATE$3:$ Move right to left by filling $1's$ such that weight of $i^{th}$ row is d
\STATE$4:$ Take transpose of the $i^{th}$ row and fill the $i^{th}$ column
\STATE$5:$ Increase $i$ by one
\STATE$6:$ Go to Step 4, if $i < n$.
\end{algorithmic}
\label{regualgo2}
\end{algorithm}

\begin{example}
The adjacency matrix for $n=6$, $d=4$, $\theta=8$, $\rho=3$ generated by Algorithm \ref{regualgo2} is
\[
A_{6 \times 6}=
\begin{bmatrix}0&1&1&1&1&0\\
1&0&0&1&1&1\\
1&0&0&1&1&1\\
1&1&1&0&0&1\\
1&1&1&0&0&1\\
0&1&1&1&1&0\\
\end{bmatrix}.
\]
\end{example}

\begin{algorithm}
\caption{Adjacency Matrix $A$ of Size $n \times n$}
\begin{algorithmic}
\REQUIRE $n , d, \theta, \rho$ and a null matrix $A$ of size $n \times n$
\ENSURE  $A_{n \times n}$ such that $weight(row[A])=d$ \newline and $weight(column[A])= \rho$
\STATE$1:$  For $1 \leq i \leq n$ and $j=n$ to $1$
\STATE$2:$ Update $A[i][j]$ and $A[j][i]$ to 1 $(i \neq j)$ such that weight of $i^{th}$ row $=d = \rho$.
\end{algorithmic}
\label{regualgo3}
\end{algorithm}

\begin{example}
The adjacency matrix for $n=6$, $d=4$, $\theta=6$, $\rho=4$ generated by Algorithm \ref{regualgo3} is
\[
A_{6 \times 6}=
\begin{bmatrix}0&0&1&1&1&1\\
0&0&1&1&1&1\\
1&1&0&0&1&1\\
1&1&0&0&1&1\\
1&1&1&1&0&0\\
1&1&1&1&0&0\\
\end{bmatrix}.
\]

\end{example}

\section{Conclusion}
In this paper, several algorithms have been presented for constructing FR codes.
Algorithm \ref{algonbytheta} is a general construction technique which for any value of $n$ calculates the possible values of $d$, $\rho$ and $\theta$,
and then generates the corresponding node-packet matrices.
The complexity of the algorithm is $\Theta(n^{3})$.
The algorithm has been tested for values up to $n=100$, and the results have been recorded.
This data is available from \url{http://www.ece.uvic.ca/~agullive/manish/List.html}.
Our aim was to generate a common data storage pattern for any given set of parameters.
The algorithm generates a node-packet matrix for each possible value of $d$, $\rho$ and $\theta$ for a range of $n$.
New algorithms were also presented for constructing regular graphs.
\section*{Acknowledgment}
The authors would like to thank Krishna Gopal Benerjee for useful discussions and Nikhil Agrawal for writing parts of the program for FR code enumeration and drawing some of the figures.


\bibliographystyle{IEEEtran}
\bibliography{cloud}

\end{document}